\documentclass[11pt,a4paper]{article}
\usepackage{jcappub,epsfig,multicol,bbm,latexsym,graphicx,subfigure}

\title{Detecting dark photon dark matter with Gaia-like astrometry observations}

\author{Huai-Ke Guo$^{a,i}$, Yingqi Ma$^b$, Jing Shu$^{a,c,d,e}$,
Xiao Xue$^{a,c}$, Qiang Yuan$^{e,f,g}$, and Yue Zhao$^h$}

\affiliation{
$^a$CAS Key Laboratory of Theoretical Physics, Insitute of Theoretical 
Physics, Chinese Academy of Sciences, Beijing 100190, P.R.China\\
$^b$National Space Science Center, Chinese Academy of Sciences, 
Beijing 100190, P.R.China\\
$^c$School of Physical Sciences, University of Chinese Academy of
Sciences, Beijing 100049, P.R.China\\
$^d$CAS Center for Excellence in Particle Physics, Beijing 100049, China\\
$^e$Center for High Energy Physics, Peking University, Beijing 100871,
P.R.China\\
$^f$Key Laboratory of Dark Matter and Space Astronomy, Purple Mountain
Observatory, Chinese Academy of Sciences, Nanjing 210008, P.R.China \\
$^g$School of Astronomy and Space Science, University of Science and
Technology of China, Hefei 230026, P.R.China\\
$^h$Department of Physics and Astronomy, University of Utah, Salt Lake 
City, UT 84112, USA
$^i$Department of Physics and Astronomy, University of Oklahoma, Norman, OK 73019, USA}
\emailAdd{ghk@itp.ac.cn}
\emailAdd{myq@nssc.ac.cn}
\emailAdd{jshu@itp.ac.cn}
\emailAdd{xuexiao@itp.ac.cn}
\emailAdd{yuanq@pmo.ac.cn}
\emailAdd{zhaoyhep@umich.edu}

\abstract{
A class of dark photon dark matter models with ultralight masses 
would lead to oscillation of a test body through a coupling with 
baryons or $B-L$ charge. This periodical oscillation of an observer
results in swing of a star's apparent position due to the effect of
aberration of light, which could be probed with high-precision
astrometry observations of stars in the Milky Way. We propose to
use the observations of stellar positions of a number of stars by 
{\tt Gaia} to search for this kind of dark photon dark matter. 
We show that this astrometry method is able to give promising 
sensitivities to search for the dark photon dark matter in the mass
range of $10^{-23}\sim10^{-21}$ eV. 
}


\begin{document}
\maketitle
\flushbottom

\section{Introduction}

About 24$\%$ of the energy density in our current Universe is composed 
by dark matter (DM). However, the identity of DM remains a big mystery. 
Very interestingly, the DM particle can be a boson with ultralight mass. 
If its mass is around $10^{-22}$ eV, its de Broglie wavelength in a dwarf 
galaxy is comparable to the core size of the galaxy. Such a DM candidate 
may provide a viable solution to the core-cusp problem in low-mass galaxies 
\cite{Hu:2000ke,Marsh:2013ywa,Bozek:2014uqa,Hui:2016ltb}.

One natural candidate for the ultralight DM particle is dark photon (DP), 
whose mass is protected by a gauge symmetry. The mass regime we are 
interested in here is extremely small, $O(10^{-22})$ eV, and the local 
occupation number for the dark photon DM (DPDM) is very large. In such a 
case, the DPDM could be treated as a background oscillating field instead 
of individual particles. 

Since we consider the scenario in which the DP mass is much smaller than 
keV, the DPDM cannot be produced thermally. One typical way for the
non-thermal production of the DPDM is through the misalignment mechanism, 
where the DM relic abundance is induced by a non-trivial initial condition 
of the dark photon field \cite{Nelson:2011sf}. More details and subtleties 
about the DPDM are further explored in \cite{Arias:2012az,Graham:2015rva}. 
Additionally, several other production mechanisms are studied recently, such as the DPDM production through the parametric resonance, the tachyonic 
instability or a network of cosmic strings
\cite{Co:2018lka,Agrawal:2018vin,Bastero-Gil:2018uel,Dror:2018pdh,Long:2019lwl}.

There have been many novel proposals to look for the DPDM in the light or 
ultralight mass regime. For example, the resonance conversion in a cavity, 
LC-circuit or dish antennas are studied in \cite{Wagner:2010mi,
Chaudhuri:2014dla,Horns:2012jf,Jaeckel:2013sqa,Knirck:2018ojz}. 
The DPDM can also be absorbed by careful choices of target materials 
which consequently causes observable excitations 
\cite{Hochberg:2016ajh,Hochberg:2016sqx,Bunting:2017net,Hochberg:2017wce,
Arvanitaki:2017nhi,Knapen:2017ekk,Baryakhtar:2018doz,Griffin:2018bjn}. 
Rather than building new experiments to look for DPDM, its existence 
can also be checked in many existing experiments, which are built for 
totally different scientific purposes 
\cite{Graham:2015ifn,Bloch:2016sjj,2018PhRvL.121f1102P,Kovetz:2018zes}.

In this paper, we focus on the scenario that the dark photon is the 
gauge boson of $U(1)_B$ or $U(1)_{B-L}$\footnote{Note that $U(1)_B$ is 
anomalous under the standard model gauge group. However, such anomaly, 
with the DP being massive, can be canceled by the Green-Schwarz anomaly 
cancellation mechanism.}. In this case, any object that carries $B$ or 
$(B-L)$ number, will experience a force caused by the DPDM background. 
This force leads to oscillations of the object, which could be consequently detected. For the very low mass range of the DPDM,
the oscillation period is about years, which is reachable by the
pulsar timing array or astrometry method. 

Here we propose to use very high-precision measurements of stellar 
locations and movements by {\tt Gaia} satellite to probe the oscillation 
effect induced by the ultralight DPDM. {\tt Gaia} is an astrometry mission
of the European Space Agency (ESA), launched in 2013 and expected to 
operate until 2022 \cite{Prusti:2016bjo}. The positions, parallaxes,
and annual proper motions of more than one billion stars with unprecedented
precision will be obtained by {\tt Gaia}, which are expected to be 
revolutionary in understanding the structure and dynamics of the Milky Way, 
the stellar physics, exoplanets, and even the fundamental particle
physics. The location accuracy of {\tt Gaia} depends on the brightness 
of the target star, and is in general about tens of micro-arcseconds 
for stars brighter than 15 mag \cite{Prusti:2016bjo}. Such an
accuracy is found to be useful in probing very tiny apparent motions
of stars caused by e.g., gravitational waves from inspirals of binary
super-massive black holes \cite{Moore:2017ity}.

\section{Effect on stellar location due to DPDM}

Let us first properly model the DPDM background. The DP is a massive 
vector boson. There are 4 components of the DPDM field, $A_\mu$, but 
only three of them  are independent. We choose to use Lorentz gauge, 
i.e. $\partial^\mu A_{\mu}=0$, in our discussion. In the non-relativistic 
limit, the Lorentz gauge implies that $A_t$ has a much smaller 
oscillation amplitude compared with spatial components $\boldsymbol{A}$. 
In addition, the contribution from $A_t$ component to the dark electric 
field is further suppressed by DM velocity, thus we will only focus on 
$\boldsymbol{A}$ in later discussions.

{Within a coherence length, $l = 2\pi/(m_A v_0)$,
the DPDM field can be approximately written as $\boldsymbol{A}
(t,\boldsymbol{x})=\boldsymbol{A_{0}}\sin(m_At-\boldsymbol{k \cdot x})$. 
Here we set the initial phase as zero without losing generality. We also 
ignore the kinetic energy contribution to the oscillation frequency. 
Typically, for $v_0 \sim 10^{-3}$, we have $l \sim 0.4
(m_A/10^{-22}~{\rm eV})^{-1}$ kpc. As mentioned before, 
we focus on the scenario where the DP is the gauge boson 
of gauged $U(1)_B$ or $U(1)_{B-L}$ group. It couples with the baryon or 
baryon-lepton charge of the test body. Analogue to a charged particle 
posed in an ordinary electromagnetic field, an acceleration is induced 
to a test mass when it carries $U(1)_B$ or $U(1)_{B-L}$ number and is 
embedded in the DPDM background. The acceleration can be approximately 
calculated as \cite{2018PhRvL.121f1102P}
\begin{equation}
\boldsymbol{a}(t,\boldsymbol{x})\simeq \epsilon e 
\frac{q}{m}m_A\boldsymbol{A_0}\cos(m_At-\boldsymbol{k \cdot x}),
\end{equation}
Here $\epsilon$ characterizes the coupling strength of the DP. It is 
normalized in terms of the electromagenetic coupling constant $e$. 
Further $q$ and $m$ are the dark charge and the mass of the test body. 
In our study, the ``charge'' $q$ equals to the total number of baryons 
or neutrons for an electric neutral object. 

One can also consider the scenario where dark photon couples to SM 
particles through kinetic mixing with the ordinary photon. In this case 
$q$ is simply the electric charge that the test body carries 
\cite{Holdom:1985ag}. However the screening effect induced by the
interstellar plasma can induce a large suppression in the parameter space 
that we can probe. Thus we will not consider this scenario in this 
study\footnote{For ordinary electromagnetic field, the interstellar 
plasma has cutoff frequency as $\omega_{\rm pe}=\sqrt{4\pi e^2n_e/m_e}
\sim 5.7\times 10^4 (n_e/{\rm cm}^{-3})^{1/2}$ Hz.
For the DPDM, the cutoff frequency is expected to be smaller by a factor 
of $\epsilon$. To see how important the screening effect can be, we first 
neglect the screening effect and estimate the sensitivity on $\epsilon$ 
that can be achieved by the method proposed in this paper. We find that 
the value of $\epsilon$ that can be probed is too large and the screening 
effects cannot be consistently ignored. Thus we will leave the discussion 
of this scenario for future studies.}.

The velocity variation due to the acceleration given in Eq. (1) is
\begin{equation}
\Delta\boldsymbol{v}(t,\boldsymbol{x})\simeq \epsilon e 
\frac{q}{m}\boldsymbol{A_0}\sin(m_At-\boldsymbol{k \cdot x}).
\end{equation}
Such a periodic velocity variation would lead to a slight swing of a 
star's apparent location, which is known as the aberration of light due 
to a moving observer. For a star with original direction $\boldsymbol{n}$
(defined as a unit vector pointing from the satellite to the star assuming
no effect from the DPDM coupling), its apparent angular deflection due 
to the velocity change $\Delta\boldsymbol{v}$ should be ($c=1$)
\begin{equation}
\Delta\theta \simeq -\Delta v~\sin\theta,
\end{equation}
where $\Delta v=|\Delta\boldsymbol{v}|$ and $\cos\theta=\Delta\boldsymbol{v}
\cdot \boldsymbol{n}/\Delta v$.

The DPDM coupling will also lead to oscillations of distant 
stars (known as the star term). The estimated angular oscillation of the 
star term is $\Delta\theta_s < \Delta v\cdot t/d\ll \Delta\theta$, where 
$t\sim1$ year is the observational time and $d\sim1$ kpc is the typical 
distance of a star\footnote{Note that we do not require the stars and the 
detector to be within one coherence length of the DPDM field.}. Only for 
stars which are very close to the Earth, e.g., $d<1$ pc, the star term 
becomes comparable to the detector-induced term eq. (2.3). Therefore we 
can neglect such oscillation effects of stars themselves.

The frequency range that {\tt Gaia} sensitive to is about
$10^{-8}\sim10^{-6}$ Hz, which corresponds to a mass range of the DPDM
of $4\times 10^{-23}\sim 4\times 10^{-21}$ eV.
For a frequency lower than $10^{-8}$ Hz, the apparent angular deflection 
varies quite slowly with time. This is similar to the proper motion of a 
star and will be largely removed when one subtracts the proper motion 
\cite{Moore:2017ity}. For a higher frequency, the observational cadence 
needs to be very high to have effective sampling of the angular deflection 
due to the DPDM. These effects will explain the behavior in the sensitivity 
estimation which will be presented in the next section.

\section{Sensitivity of DPDM searches with Gaia}

To perform a solid estimation on the sensitivity that can be achieved 
by our proposed analysis using {\tt Gaia}-like astrometry observations, 
we simulate stellar motions with and without the DPDM coupling. Following 
Ref. \cite{Moore:2017ity}, we assume that the orbital motion of 
the satellite surrounding the Earth and Sun can be precisely corrected,
leaving only the stellar proper motion and the DPDM coupling effects 
to be considered here. This is reasonable for {\tt Gaia} which employs
a series of orbit control and reconstruction techniques to ensure an 
accuracy of tens of $\mu$as of the stellar location measurement 
\cite{Prusti:2016bjo}. 

We use a quadratic model to approximately describe the proper motion of 
a star. For each star, its initial velocity and acceleration are randomly 
assigned, assuming Gaussian distributions with mean values as zero and 
standard deviations as 50 km~s$^{-1}$ and 20 km~s$^{-1}$~yr$^{-1}$, for 
the right ascension and declination directions. The proper motion can 
be subtracted through a fit to multiple measurements, $\sim O(100)$, 
within the lifetime of the mission \cite{Moore:2017ity}. 

The angular deflection induced by the DPDM 
is further added on top of the proper motion. The direction of the 
gauge field $\boldsymbol{A}$ is described by an equatorial coordinate
$(\alpha, \delta)$ which are free parameters to be inferred from the
data. From Eq.~(2), DPDM induced change in velocity is written as
\begin{equation}
\Delta\boldsymbol{v}(t,\boldsymbol{x})\simeq \epsilon e 
\frac{q}{m}A_0\boldsymbol{n_0}\sin[m_A(t-t_0)+\phi],
\end{equation}
where $\boldsymbol{n_0}=(\cos\delta\cos\alpha,\,\cos\delta\sin\alpha,\,
\sin\delta)$, $t_0$ is the zero point of the simulated observation.
Here we do not include the phase change caused by 
$\boldsymbol {k \cdot x}$, which is safely negligible in the parameter 
region we are interested in. For example, the coherence length for the 
parameter space we are interested in is about $0.01 \sim 1$ kpc given 
the velocity $v_0\sim 10^{-3}$. The largest motion of the satellite in 
the Milky Way is about $3\times10^{-4}$~pc~yr$^{-1}$, assuming again a 
velocity of $\sim10^{-3}$ of the Sun. The movement distance of the 
detector is thus significantly smaller than the coherence length of 
the DPDM, and it is safe to treat the DPDM field as spatially universal in the local environment.

We randomly generate $10^4$ stars uniformly distributed in the 
sky.\footnote{This is a zeroth-order approximation. Considering a more
realistic distribution of stars with a concentration along the Galactic
disk will result in a non-uniform sensitivity for the DPDM field with
different orientation. Including this subtlety does not change our 
qualitative conclusion.} For each star, 75 observations are performed 
within 5 years with a uniform cadence, which is comparable to the
average design performance of {\tt Gaia} at the end of mission 
\cite{Prusti:2016bjo}. The localization accuracy of the {\tt Gaia} 
satellite is assumed to be $100~\mu$as. Note that in reality such an 
accuracy depends on the magnitude of a star \cite{Prusti:2016bjo}. 
This subtlety is not included in the current study. We adopt the data 
compression technique proposed in Ref.~\cite{Moore:2017ity}, 
which gives a significantly improved
location accuracy by measurements of a large number of 
stars in a small sky cell. Here we assume a compression ratio of 
$10^9/10^4=10^5$, which gives a noise level of $\sigma=100~\mu{\rm as}/
\sqrt{10^5}$. In our study, for each observation of a star, a Gaussian 
noise with $\sigma$ is added when we simulate the positions of stars. 
The proper motion parameters (in total 4 parameters describing the
velocity and acceleration in two orthogonal directions)  are derived
through fitting the 75 observations of each star with the quadratic
model. At last, we subtract the proper motion with the best-fit 
parameters in order to study the remaining motion of stars. 

As an illustration, we assume that the dark photon is the gauge boson 
of $U(1)_B$, and we take the input parameters as 
$(m_A,\epsilon,\phi,\alpha,\delta)=(10^{-22}~{\rm eV}, 3\times10^{-24}, 
2.59, 1.25, 0.68)$. Except for the mass and the coupling
constant, the other parameters are randomly generated. 
 We employ the 
Markov Chain Monte Carlo (MCMC) method \cite{Lewis:2002ah} to fit the 
model parameters from the mock data. We find that the model with the 
DPDM interaction gives significantly better fit to the simulation data. 
Compared with the {\tt null} hypothesis without the DPDM interaction, 
the $\chi^2$ value of the model with the DPDM interaction is smaller by 
about $70$. The fitting distributions of the model parameters are shown 
in Figure~\ref{fig:triangle}, from which we can see that the input 
parameters are reasonably reproduced.

\begin{figure}[!htb]
\centering
\includegraphics[width=0.7\textwidth]{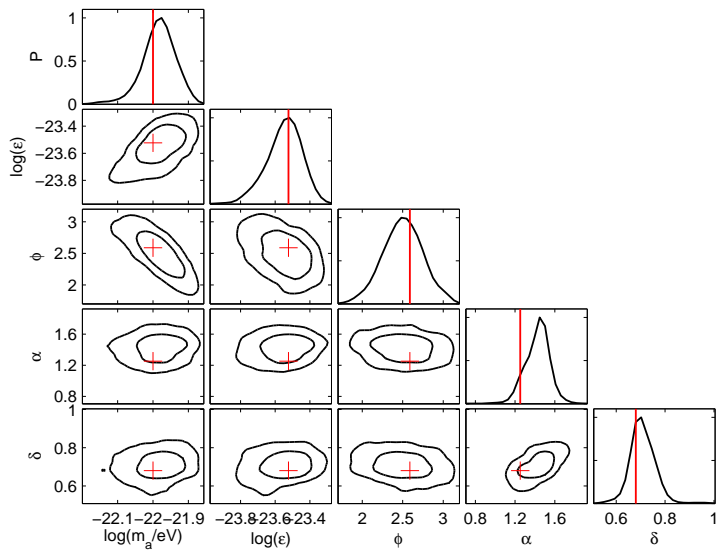}
\caption{Fitting 1-dimensional probability distributions (diagonal) 
and 2-dimensional credible regions (off-diagonal; $68\%$ and $95\%$ 
credible levels from inside to outside) of the model parameters based 
on simulated astrometry data of $10^4$ stars. The red lines and crosses
show the input parameters of the simulation.
\label{fig:triangle}}
\end{figure}

To estimate the sensitivity of detecting the DPDM that the astrometry 
observations can reach, we vary the coupling constant $\epsilon$ for
different choices of $m_A$, and repeat the above simulation and
analysis. The $95\%$ confidence level of a detection is 
defined as that the difference of the $\chi^2$ between the
{\tt signal} hypothesis and the {\tt null} hypothesis is about $9.7$, 
with four additional degrees of freedom in the {\tt signal} hypothesis.

\begin{figure}[!htb]
\centering
\includegraphics[width=0.48\textwidth]{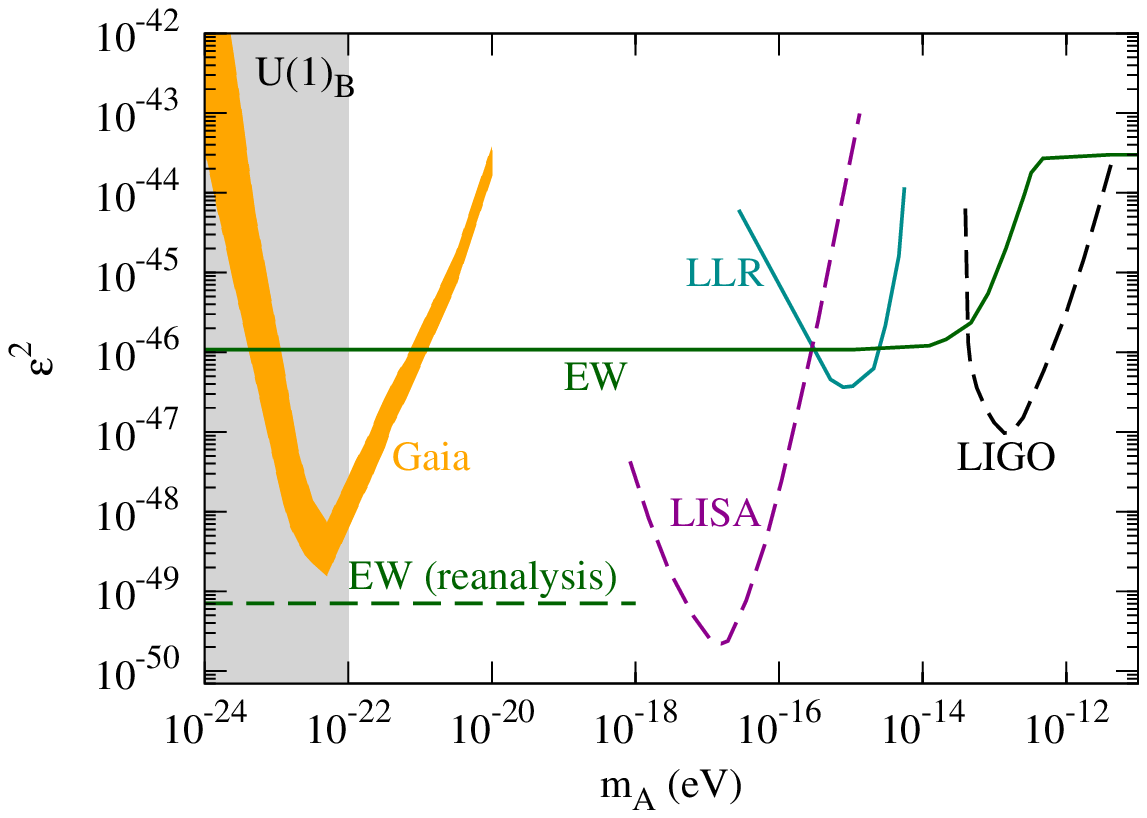}
\includegraphics[width=0.48\textwidth]{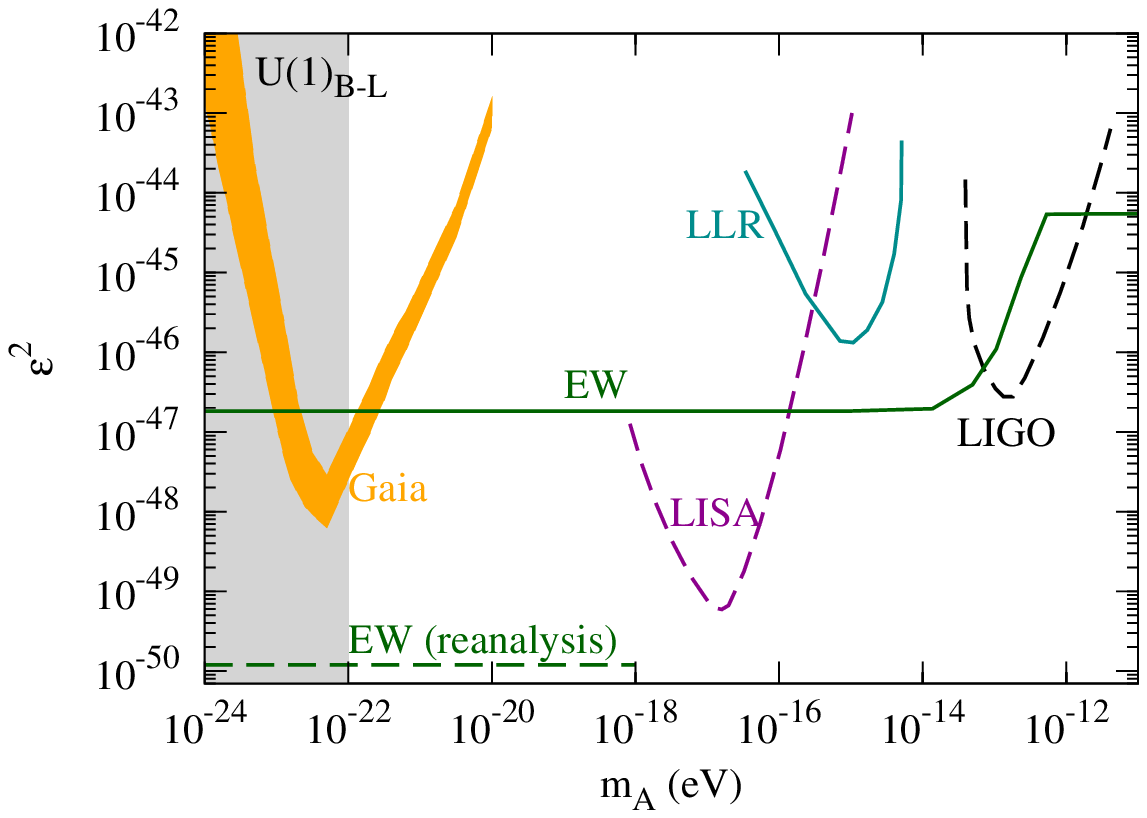}
\caption{Expected $2\sigma$ exclusion limits of the DPDM coupling constant 
from 5-year astrometry observations of {\tt Gaia} (orange bands), for the
$U(1)_B$ (left) and $U(1)_{B-L}$ (right) types of coupling, compared with 
those from the EW experiments \cite{Su:1994gu,Schlamminger:2007ht,
Graham:2015ifn}, the LLR experiment \cite{Talmadge:1988qz,
Williams:2004qba,Turyshev:2006gm}, and the expected sensitivities 
of gravitational wave experiments LISA and LIGO \cite{2018PhRvL.121f1102P}.
The grey shaded region marks out the parameter region with $m_A<10^{-22}$ 
eV, in which the de Broglie wavelength of the DPDM is larger than the 
typical size of a dwarf galaxy \cite{Hu:2000ke}.
\label{fig:limit}}
\end{figure}

The derived sensitivities of the coupling constant $\epsilon^2$ as 
functions of the DPDM mass are shown in Figure~\ref{fig:limit}, for the
$U(1)_B$ (top) and $U(1)_{B-L}$ (bottom) types of coupling. Note
that the sensitivities are presented by a band which is the envelope
of the results considering fluctuations due to different realizations
of the sample and the random choices of the initial phase. 

We note that if $m_A$ is smaller than $10^{-22.5}$ eV, the oscillation
frequency becomes very small, and the signal starts to degenerate with
the proper motion for a limited observation time (a few years). 
In this case the sensitivities become weaker. Also we can see that
the sensitivity band becomes wider when $m_A$ is smaller. This is because 
in the low frequency region, the signal degenerates with the background 
to different extent for different initial phase of the DPDM field, and 
the background subtraction gives different results.

For comparison, the experimental limits from the Eot-Wash (EW) experiments 
\cite{Su:1994gu,Schlamminger:2007ht,Graham:2015ifn} and the Lunar 
Laser Ranging (LLR) experiment \cite{Talmadge:1988qz,Williams:2004qba,
Turyshev:2006gm}, and the expected sensitivities from the gravitational 
wave experiments LISA and LIGO \cite{2018PhRvL.121f1102P} are also shown.
In Ref. \cite{Graham:2015ifn} a re-scaling of the EW experiment result of
\cite{Schlamminger:2007ht} was performed to convert the static force signal
induced by the Earth to the DM-induced signal, which is labelled as
``EW (reanalysis)''. However, a dedicated reanalysis may be necessary to 
give actual bounds on the parameter space. The astrometry method proposed
in this work shows promising sensitivities for the DPDM mass range of 
$m_A\lesssim10^{-21}$ eV, corresponding to frequencies smaller than 
$10^{-7}$ Hz.

\section{Conclusion}

In this work we propose to use the high-precision astrometry measurements, 
i.e. {\tt Gaia}-like satellite, to search for the ultralight DM candidate. 
We consider the scenario where DM is composed by dark photon, which is the 
gauge boson of $U(1)_B$ or $U(1)_{B-L}$. In such scenario, the existence 
of DPDM is expected to lead a periodic oscillation of the satellite. 
This results in angular deflections of target stars due to the aberration 
effect. Benefiting from precise location measurements of a large number of 
stars, even a very weak DPDM coupling can potentially be revealed by extracting a universal oscillation pattern of all stars. 
We find that this proposed search strategy can probe a large unexplored 
parameter space of DPDM, e.g. a coupling as small as $\epsilon\sim10^{-24}$ 
in the mass range of $10^{-23}\sim10^{-21}$ eV.

There are other kinds of oscillation effects induced by various types
of ultralight DM candidates, e.g., the axion, can also be probed with
the astrometry method. The method studied in the paper serves an important 
complement in probing a class of ultralight DM models, which is
proposed to be detected by the pulsar timing array 
\cite{Khmelnitsky:2013lxt,DeMartino:2017qsa,Porayko:2018sfa}.

\acknowledgments
J.S. is supported by the National Natural Science Foundation of China 
(NSFC) under grant No.11847612, No.11690022, No.11851302, No.11675243 
and No.11761141011, and also supported by the Strategic Priority Research 
Program of the Chinese Academy of Sciences under grant No.XDB21010200 
and No.XDB23000000. 
Q.Y. is supported by the National Key Research and Development Program 
of China under grant No.2016YFA0400204, the NSFC under grant No.11722328, 
No.11851305, and the 100 Talents program of Chinese Academy of Sciences.


\begin{thebibliography}{99}


\bibitem{Hu:2000ke} 
  W.~Hu, R.~Barkana and A.~Gruzinov,
  Cold and fuzzy dark matter,
  Phys.\ Rev.\ Lett.\  {\bf 85}, 1158 (2000)
  [astro-ph/0003365].

\bibitem{Marsh:2013ywa} 
  D.~J.~E.~Marsh and J.~Silk,
  A Model For Halo Formation With Axion Mixed Dark Matter,
  Mon.\ Not.\ Roy.\ Astron.\ Soc.\  {\bf 437}, no. 3, 2652 (2014)
  [arXiv:1307.1705 [astro-ph.CO]].

\bibitem{Bozek:2014uqa} 
  B.~Bozek, D.~J.~E.~Marsh, J.~Silk and R.~F.~G.~Wyse,
  Galaxy UV-luminosity function and reionization constraints on axion dark matter,
  Mon.\ Not.\ Roy.\ Astron.\ Soc.\  {\bf 450}, no. 1, 209 (2015)
  [arXiv:1409.3544 [astro-ph.CO]].

\bibitem{Hui:2016ltb} 
  L.~Hui, J.~P.~Ostriker, S.~Tremaine and E.~Witten,
  Ultralight scalars as cosmological dark matter,
  Phys.\ Rev.\ D {\bf 95}, no. 4, 043541 (2017)
  [arXiv:1610.08297 [astro-ph.CO]].


\bibitem{Nelson:2011sf} 
  A.~E.~Nelson and J.~Scholtz,
  Dark Light, Dark Matter and the Misalignment Mechanism,
  Phys.\ Rev.\ D {\bf 84}, 103501 (2011)
  [arXiv:1105.2812 [hep-ph]].

\bibitem{Arias:2012az} 
  P.~Arias, D.~Cadamuro, M.~Goodsell, J.~Jaeckel, J.~Redondo and A.~Ringwald,
  WISPy Cold Dark Matter,
  JCAP {\bf 1206}, 013 (2012)
  [arXiv:1201.5902 [hep-ph]].

\bibitem{Graham:2015rva} 
  P.~W.~Graham, J.~Mardon and S.~Rajendran,
  Vector Dark Matter from Inflationary Fluctuations,
  Phys.\ Rev.\ D {\bf 93}, no. 10, 103520 (2016)
  [arXiv:1504.02102 [hep-ph]].

\bibitem{Co:2018lka} 
  R.~T.~Co, A.~Pierce, Z.~Zhang and Y.~Zhao,
  Dark Photon Dark Matter Produced by Axion Oscillations,
  arXiv:1810.07196 [hep-ph].

\bibitem{Agrawal:2018vin} 
  P.~Agrawal, N.~Kitajima, M.~Reece, T.~Sekiguchi and F.~Takahashi,
  Relic Abundance of Dark Photon Dark Matter,
  arXiv:1810.07188 [hep-ph].

\bibitem{Bastero-Gil:2018uel} 
  M.~Bastero-Gil, J.~Santiago, L.~Ubaldi and R.~Vega-Morales,
  Vector dark matter production at the end of inflation,
  arXiv:1810.07208 [hep-ph].

\bibitem{Dror:2018pdh} 
  J.~A.~Dror, K.~Harigaya and V.~Narayan,
  Parametric Resonance Production of Ultralight Vector Dark Matter,
  arXiv:1810.07195 [hep-ph].

\bibitem{Long:2019lwl} 
  A.~J.~Long and L.~T.~Wang,
  Dark Photon Dark Matter from a Network of Cosmic Strings,
  arXiv:1901.03312 [hep-ph].

\bibitem{Wagner:2010mi} 
  A.~Wagner {\it et al.} [ADMX Collaboration],
  A Search for Hidden Sector Photons with ADMX,
  Phys.\ Rev.\ Lett.\  {\bf 105}, 171801 (2010)
  [arXiv:1007.3766 [hep-ex]].

\bibitem{Chaudhuri:2014dla} 
  S.~Chaudhuri, P.~W.~Graham, K.~Irwin, J.~Mardon, S.~Rajendran and Y.~Zhao,
  Radio for hidden-photon dark matter detection,
  Phys.\ Rev.\ D {\bf 92}, no. 7, 075012 (2015)
  [arXiv:1411.7382 [hep-ph]].

\bibitem{Horns:2012jf} 
  D.~Horns, J.~Jaeckel, A.~Lindner, A.~Lobanov, J.~Redondo and A.~Ringwald,
  Searching for WISPy Cold Dark Matter with a Dish Antenna,
  JCAP {\bf 1304}, 016 (2013)
  [arXiv:1212.2970 [hep-ph]].

\bibitem{Jaeckel:2013sqa} 
  J.~Jaeckel and J.~Redondo,
  An  antenna for directional detection of WISPy dark matter,
  JCAP {\bf 1311}, 016 (2013)
  [arXiv:1307.7181 [hep-ph]].


\bibitem{Knirck:2018ojz} 
  S.~Knirck, T.~Yamazaki, Y.~Okesaku, S.~Asai, T.~Idehara and T.~Inada,
  First results from a hidden photon dark matter search in the meV sector using a plane-parabolic mirror system,
  JCAP {\bf 1811}, no. 11, 031 (2018)
  [arXiv:1806.05120 [hep-ex]].

\bibitem{Hochberg:2016ajh} 
  Y.~Hochberg, T.~Lin and K.~M.~Zurek,
  Detecting Ultralight Bosonic Dark Matter via Absorption in Superconductors,
  Phys.\ Rev.\ D {\bf 94}, no. 1, 015019 (2016)
  [arXiv:1604.06800 [hep-ph]].

\bibitem{Hochberg:2016sqx} 
  Y.~Hochberg, T.~Lin and K.~M.~Zurek,
  Absorption of light dark matter in semiconductors,
  Phys.\ Rev.\ D {\bf 95}, no. 2, 023013 (2017)
  [arXiv:1608.01994 [hep-ph]].

\bibitem{Bunting:2017net} 
  P.~C.~Bunting, G.~Gratta, T.~Melia and S.~Rajendran,
  Magnetic Bubble Chambers and Sub-GeV Dark Matter Direct Detection,
  Phys.\ Rev.\ D {\bf 95}, no. 9, 095001 (2017)
  [arXiv:1701.06566 [hep-ph]].

\bibitem{Hochberg:2017wce} 
  Y.~Hochberg {\it et al.},
  Detection of sub-MeV Dark Matter with Three-Dimensional Dirac Materials,
  Phys.\ Rev.\ D {\bf 97}, no. 1, 015004 (2018)
  [arXiv:1708.08929 [hep-ph]].

\bibitem{Arvanitaki:2017nhi} 
  A.~Arvanitaki, S.~Dimopoulos and K.~Van Tilburg,
  Resonant absorption of bosonic dark matter in molecules,
  Phys.\ Rev.\ X {\bf 8}, no. 4, 041001 (2018)
  [arXiv:1709.05354 [hep-ph]].

\bibitem{Knapen:2017ekk} 
  S.~Knapen, T.~Lin, M.~Pyle and K.~M.~Zurek,
  Detection of Light Dark Matter With Optical Phonons in Polar Materials,
  Phys.\ Lett.\ B {\bf 785}, 386 (2018)
  [arXiv:1712.06598 [hep-ph]].

\bibitem{Baryakhtar:2018doz} 
  M.~Baryakhtar, J.~Huang and R.~Lasenby,
  Axion and hidden photon dark matter detection with multilayer optical haloscopes,
  Phys.\ Rev.\ D {\bf 98}, no. 3, 035006 (2018)
  [arXiv:1803.11455 [hep-ph]].

\bibitem{Griffin:2018bjn} 
  S.~Griffin, S.~Knapen, T.~Lin and K.~M.~Zurek,
  Directional Detection of Light Dark Matter with Polar Materials,
  Phys.\ Rev.\ D {\bf 98}, no. 11, 115034 (2018)
  [arXiv:1807.10291 [hep-ph]].



\bibitem{Graham:2015ifn} 
  P.~W.~Graham, D.~E.~Kaplan, J.~Mardon, S.~Rajendran and W.~A.~Terrano,
  Dark Matter Direct Detection with Accelerometers,
  Phys.\ Rev.\ D {\bf 93}, no. 7, 075029 (2016)
  [arXiv:1512.06165 [hep-ph]].

\bibitem{Bloch:2016sjj} 
  I.~M.~Bloch, R.~Essig, K.~Tobioka, T.~Volansky and T.~T.~Yu,
  Searching for Dark Absorption with Direct Detection Experiments,
  JHEP {\bf 1706}, 087 (2017)
  [arXiv:1608.02123 [hep-ph]].

\bibitem{2018PhRvL.121f1102P} 
  A.~Pierce, K.~Riles and Y.~Zhao,
  Searching for Dark Photon Dark Matter with Gravitational Wave Detectors,
  Phys.\ Rev.\ Lett.\  {\bf 121}, no. 6, 061102 (2018)
  [arXiv:1801.10161 [hep-ph]].

\bibitem{Kovetz:2018zes} 
  E.~D.~Kovetz, I.~Cholis and D.~E.~Kaplan,
  Bounds on Ultra-Light Hidden-Photon Dark Matter from 21cm at Cosmic Dawn,
  arXiv:1809.01139 [astro-ph.CO].

\bibitem{Prusti:2016bjo} 
  T.~Prusti {\it et al.} [Gaia Collaboration],
  The Gaia Mission,
  Astron.\ Astrophys.\  {\bf 595}, A1 (2016)
  [arXiv:1609.04153 [astro-ph.IM]].

\bibitem{Moore:2017ity} 
  C.~J.~Moore, D.~Mihaylov, A.~Lasenby and G.~Gilmore,
  Astrometric Search Method for Individually Resolvable Gravitational Wave Sources with Gaia,
  Phys.\ Rev.\ Lett.\  {\bf 119}, no. 26, 261102 (2017)
  doi:10.1103/PhysRevLett.119.261102
  [arXiv:1707.06239 [astro-ph.IM]].

\bibitem{Holdom:1985ag} 
  B.~Holdom,
  Two U(1)'s and Epsilon Charge Shifts,
  Phys.\ Lett.\  {\bf 166B}, 196 (1986).

\bibitem{Lewis:2002ah} 
  A.~Lewis and S.~Bridle,
  Cosmological parameters from CMB and other data: A Monte Carlo approach,
  Phys.\ Rev.\ D {\bf 66}, 103511 (2002)
  [astro-ph/0205436].

\bibitem{Su:1994gu} 
  Y.~Su, B.~R.~Heckel, E.~G.~Adelberger, J.~H.~Gundlach, M.~Harris, G.~L.~Smith and H.~E.~Swanson,
  New tests of the universality of free fall,
  Phys.\ Rev.\ D {\bf 50}, 3614 (1994).

\bibitem{Schlamminger:2007ht} 
  S.~Schlamminger, K.-Y.~Choi, T.~A.~Wagner, J.~H.~Gundlach and E.~G.~Adelberger,
  Test of the equivalence principle using a rotating torsion balance,
  Phys.\ Rev.\ Lett.\  {\bf 100}, 041101 (2008)
  [arXiv:0712.0607 [gr-qc]].

\bibitem{Talmadge:1988qz} 
  C.~Talmadge, J.~P.~Berthias, R.~W.~Hellings and E.~M.~Standish,
  Model Independent Constraints on Possible Modifications of Newtonian Gravity,
  Phys.\ Rev.\ Lett.\  {\bf 61}, 1159 (1988).

\bibitem{Williams:2004qba} 
  J.~G.~Williams, S.~G.~Turyshev and D.~H.~Boggs,
  Progress in lunar laser ranging tests of relativistic gravity,
  Phys.\ Rev.\ Lett.\  {\bf 93}, 261101 (2004)
  [gr-qc/0411113].

\bibitem{Turyshev:2006gm} 
  S.~G.~Turyshev and J.~G.~Williams,
  Space-based tests of gravity with laser ranging,
  Int.\ J.\ Mod.\ Phys.\ D {\bf 16}, 2165 (2007)
  [gr-qc/0611095].

\bibitem{Khmelnitsky:2013lxt} 
  A.~Khmelnitsky and V.~Rubakov,
  Pulsar timing signal from ultralight scalar dark matter,
  JCAP {\bf 1402}, 019 (2014)
  [arXiv:1309.5888 [astro-ph.CO]].

\bibitem{DeMartino:2017qsa} 
  I.~De Martino, T.~Broadhurst, S.~H.~Henry Tye, T.~Chiueh, H.~Y.~Schive and R.~Lazkoz,
  Recognizing Axionic Dark Matter by Compton and de Broglie Scale Modulation of Pulsar Timing,
  Phys.\ Rev.\ Lett.\  {\bf 119}, no. 22, 221103 (2017)
  [arXiv:1705.04367 [astro-ph.CO]].

\bibitem{Porayko:2018sfa} 
  N.~K.~Porayko {\it et al.},
  Parkes Pulsar Timing Array constraints on ultralight scalar-field dark matter,
  Phys.\ Rev.\ D {\bf 98}, no. 10, 102002 (2018)
  [arXiv:1810.03227 [astro-ph.CO]].

\end{thebibliography}
\end{document}